\documentclass[epj,final]{svjour}
\usepackage{graphics}
\usepackage{float}
\usepackage{braket}
\usepackage{subfigure}
\usepackage{lipsum}

\usepackage{hyperref}
\hypersetup{
    colorlinks=true,
    linkcolor=blue,
    citecolor=blue
}

\begin{document}
\title{Quantum correlations between two distant cavity QED systems coupled by a mechanical resonator}
\author{J. E. Ram\'irez-Mu\~noz \thanks{\emph{e-mail:} jeramirezm@unal.edu.co} \and J. P. Restrepo Cuartas \and H. Vinck-Posada
%
}                     
\offprints{}          
\institute{Universidad Nacional de Colombia - Sede Bogot\'a, Facultad de Ciencias, Departamento de F\'isica\\
 Carrera 45 No. 26-85, C.P. 111321, Bogot\'a, Colombia
}
\date{Received: date / Revised version: date}
%
\abstract{
Achieving quantum correlations between two distant systems is a desirable feature for quantum networking. In this work, we study a system composed of two quantum emitter-cavity subsystems spatially separated. A mechanical resonator couples to either both quantum emitters or both cavities leading to quantum correlations between both subsystems such as non-local light-matter dressed states and cavity-cavity normal mode splitting. These indirect couplings can be explained by an effective Hamiltonian for large energy detuning between the mechanical resonator and the atoms/cavities. Moreover, it is found optimal conditions for the physical parameters of the system in order to maximize the entanglement of such phonon-mediated couplings.
%
} 
\maketitle
\section{Introduction}
\label{intro}
Photonic channels are the most common in standard quantum networking \cite{Kimble2008,Hammerer2010}. However, other mechanisms can be envisaged to perform the quantum information processing tasks, for example, phononic channels. On one hand, cavity quantum electrodynamics systems (cQED) have offered great potentialities in quantum computing \cite{Nielsen2000,Monroe2002}, especially, in semiconductor nanoestructures \cite{Lodahl2015,Laussy2007a} and superconducting circuits platforms \cite{Wallraff2004j,Devoret2013,You2005,You2011}. On the other hand, in the cavity optomechanics frame (COM) \cite{Aspelmeyer2014}, mechanical resonators coupled to cavities and artificial atoms allow controlling and enhancing quantum properties at the same time that introduce mechanics in the quantum realm \cite{Schwab2005}. Hybrid optomechanical systems \cite{Xiang2013,Pirkkalainen2013,Pirkkalainen2015}, involving both cQED and COM, would provide astonishing opportunities for quantum networking \cite{Dong2015}. 

Entanglement as the main resource for quantum computation \cite{Horodecki2009} is aimed to link a whole quantum network composed of atoms trapped in optical cavities (nodes) linked by photons propagating from one to others (channels). Particularly, entangling two quantum nodes of a network in a reversible way is a required condition to distribute entanglement across the network and teleport quantum states. In such multipartite systems it is worth to study which parts or subsystems are most entangled than others and how to improve that entanglement \cite{Liao2018,Yang2017}. Now, phononic modes of mechanical resonators could be considered to connect cQED nodes as an alternative to the standard photonic channels. Most of the works so far consider that phononic modes modulate the energies of quantum emitters and cavities \cite{Restrepo2017}. Beyond the dispersive regime, other coupling mechanisms have been explored such as linear coupling \cite{Ramirez-Munoz2018} or by mechanical variation of the Rabi coupling rate \cite{Cotrufo2017,Hammerer2009}. In this work, we study a double quantum emitter-cavity system. Both subsystems are coupled linearly by a single mechanical mode of a mechanical resonator. This phononic mode couples either the quantum emitters or the cavities. 

The rest of the paper is organized as follows, in Sect. \ref{sec2} we set the theoretical model for the mechanically coupled cavity QED subsystems. Then, in Sect. \ref{sec3} we study the Hamiltonian by numerical diagonalization and derive an effective Hamiltonian in the dispersive regime which gives accounts of the main effective couplings between the different parts of the system. Besides, we quantify the bipartite entanglement of such dressed states. Finally, in Sect. \ref{conclusions}, we discuss and conclude.

\section{Theoretical framework}
\label{sec2}

\begin{figure}[H]
\centering
\resizebox{0.45\textwidth}{!}{%
	\includegraphics{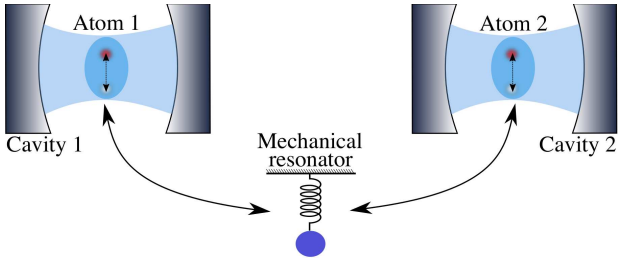}
}
\caption{Sketch of the double cavity QED system studied. The quantum emitters couple to the same mechanical mode.}
\label{system}
\end{figure}

The system considered here is schematically depicted in Fig. \ref{system} and consist of two distant quantum emitter-cavity systems each one interacting via dipole interaction. The quantum emitter is considered as a two-level system (TLS). A single-mode of a mechanical resonator interacts with each quantum emitter but not with the cavities.

We assume strong coupling between quantum emitters and cavities such that each subsystem is modeled with the Jaynes-Cummings Hamiltonian:

\begin{equation}
\hat{H}_{1}=\omega_{c1}\hat{a}_{1}^{\dagger}\hat{a}_{1}+\omega_{a1}\hat{\sigma}_{1}^{\dagger}\hat{\sigma}_{1}+g_{1}\left(\hat{a}_{1}^{\dagger}\hat{\sigma}_{1}+\hat{a}_{1}\hat{\sigma}_{1}^{\dagger}\right)
\end{equation}
and
\begin{equation}
\hat{H}_{2}=\omega_{c2}\hat{a}_{2}^{\dagger}\hat{a}_{2}+\omega_{a2}\hat{\sigma}_{2}^{\dagger}\hat{\sigma}_{2}+g_{2}\left(\hat{a}_{2}^{\dagger}\hat{\sigma}_{2}+\hat{a}_{2}\hat{\sigma}_{2}^{\dagger}\right)
\end{equation}
where $\omega_{c1}$ and $\omega_{c2}$ are the cavity energies, $\omega_{a1}$ and $\omega_{a2}$ are the atom energies and, $g_{1}$ and $g_{2}$ are the light-matter interaction strengths in each subsystem. 

As for the mechanical mode, we consider linear coupling with each quantum emitter as it has been proposed in some works before \cite{Cotrufo2017,Hammerer2009}:

\begin{eqnarray}
\hat{H}_{m}=\omega_{m}\hat{b}^{\dagger}\hat{b}+g_{m1}\left(\hat{b}^{\dagger}\hat{\sigma}_{1}+\hat{b}\hat{\sigma}_{1}^{\dagger} \right)+g_{m2}\left(\hat{b}^{\dagger}\hat{\sigma}_{2}+\hat{b}\hat{\sigma}_{2}^{\dagger} \right)
\end{eqnarray}
here $\omega_{m}$ is the energy of the phonon mode of the mechanical resonator and, $g_{m1}$ and $g_{m2}$ are the coupling strength rates between the mechanical mode and each quantum emitter. The total Hamiltonian is then:

\begin{equation}
\hat{H}=\hat{H}_{1}+\hat{H}_{2}+\hat{H}_{m}
\label{total_hamiltonian}
\end{equation}

Since the Hamiltonian commutes with the total number operator ($\hat{N}=\hat{a}_{1}^{\dagger}\hat{a}_{1}+\hat{a}_{2}^{\dagger}\hat{a}_{2}+\hat{\sigma}_{1}^{\dagger}\hat{\sigma}_{1}+\hat{\sigma}_{2}^{\dagger}\hat{\sigma}_{2}+\hat{b}^{\dagger}\hat{b}$), then it can be diagonalized for each excitation manifold composed of all states $\ket{\alpha,n,\beta,m,\ell}$ with $\alpha+n+\beta+m+\ell=\textnormal{constant}$. Throughout the work, the notation for the states is as follows: $\ket{\textnormal{Atom1,Cav1,Atom2,Cav2,phonon}}$.\\

With regard to the physical parameters, we approach the problem not with absolute values but rather with ratios between them in order to find effects for a great variety of systems that can satisfy the conditions studied throughout the work. Besides, we consider $g_{m1}=g_{m2}=g_{m}$ and $g_{1}=g_{2}$. Particularly, we explore situations where atomic and photonic frequencies are not much larger than the mechanical interaction rates but still are out of resonance with the mechanical resonator such that a large detuning approximation can be addressed. These requirements are not far from experimental works, for example, in the context of circuit QED \cite{Pirkkalainen2013,Pirkkalainen2015}, parameter ratios have been achieved as follows $\omega_{c(a)}/g_{m}\approx 190$, $\omega_{c}/\omega_{m}\approx 67$ and $\omega_{m}/g_{m}\approx 3$.

\section{Dressed states and entanglement}
\label{sec3}

\subsection{Inter-cavity normal mode splittings}

\begin{figure*}
\centering
\resizebox{0.45\textwidth}{!}{%
\subfigure{\includegraphics{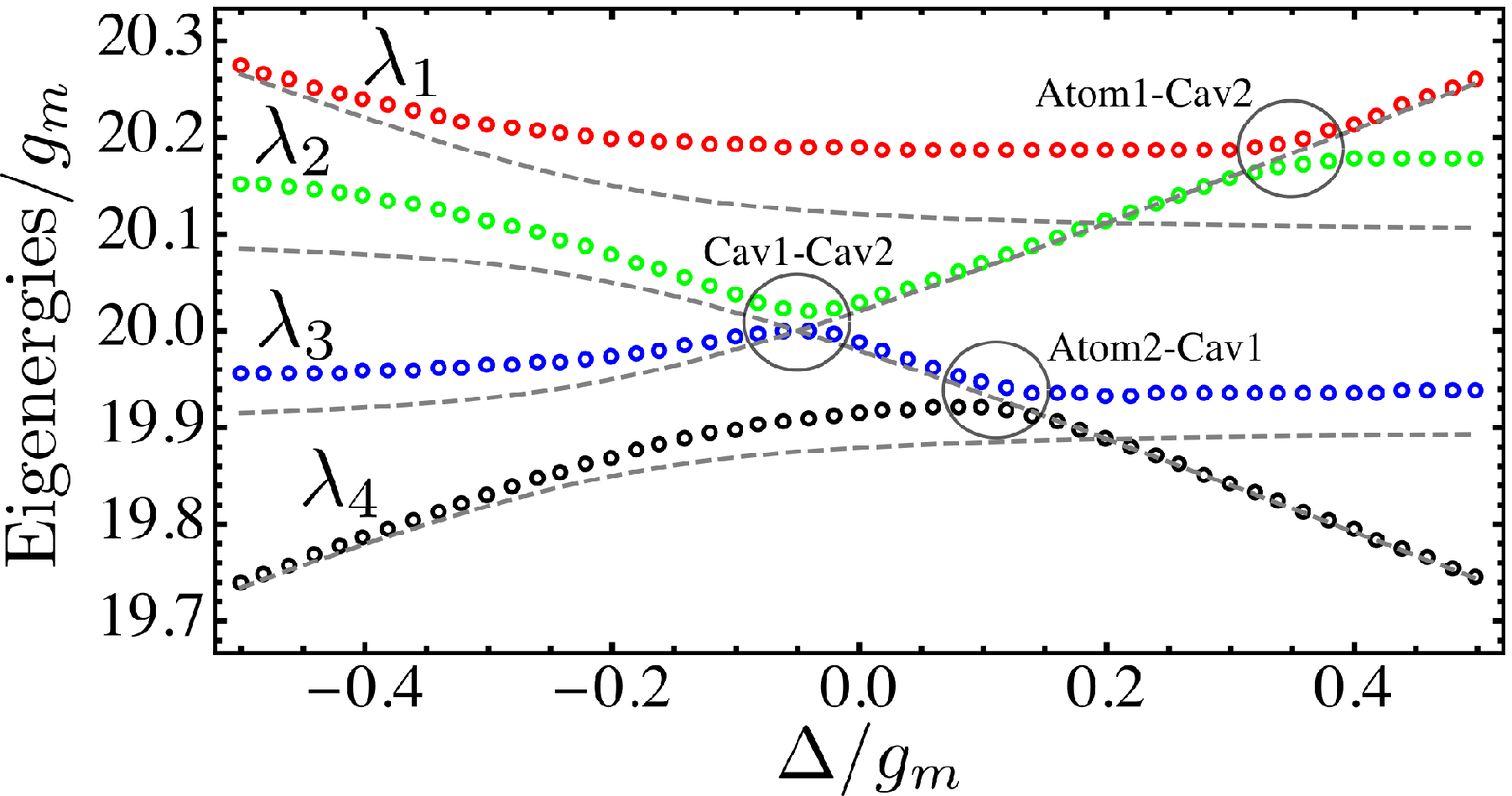}}
}\hspace{0.5cm}
\resizebox{0.45\textwidth}{!}{%
\subfigure{\includegraphics{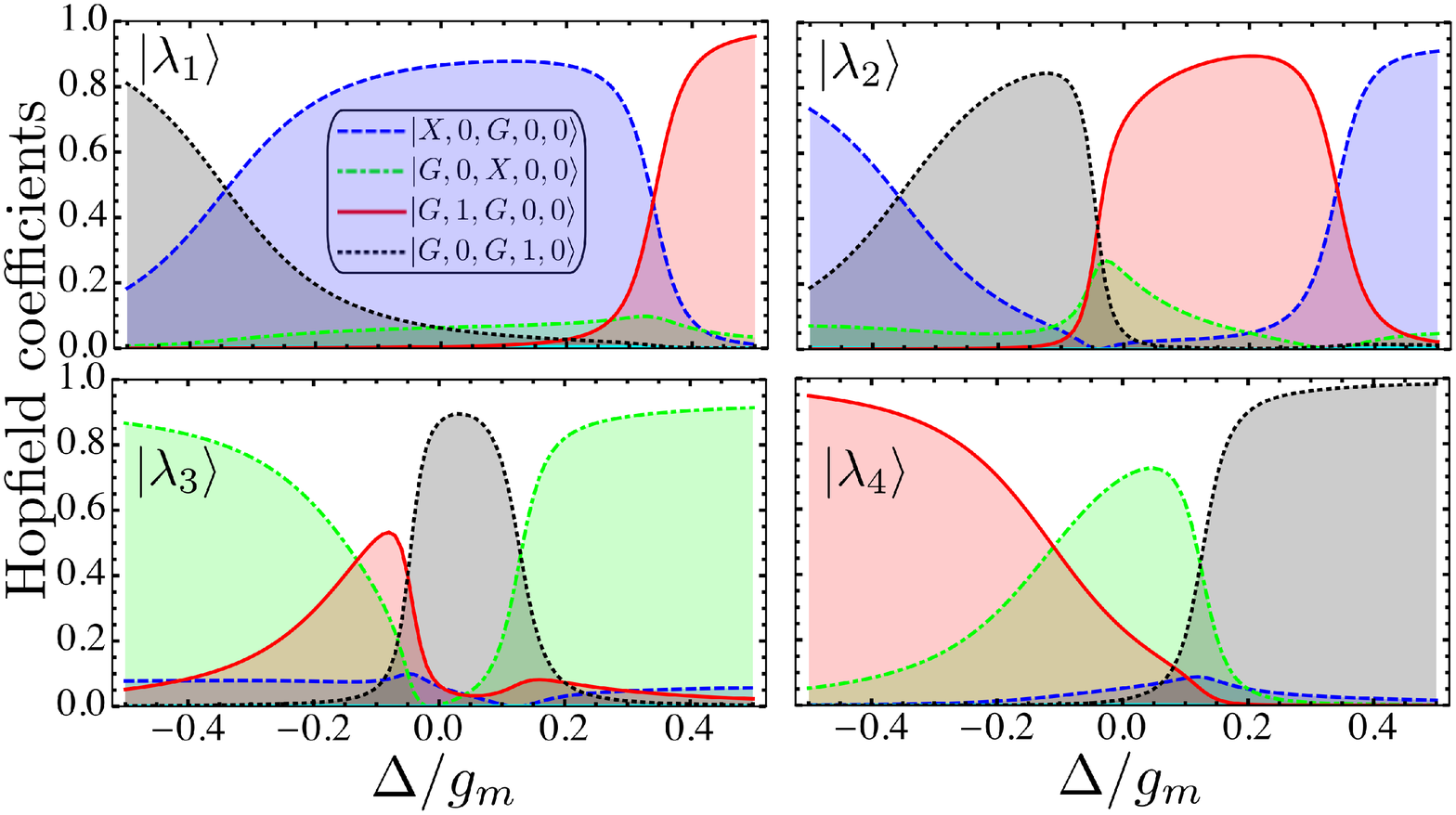}}
}
\caption{Dressed states. Left plot: Eigenenergies for the first excitation manifold as a function of the inter-cavity energy detuning, $\Delta$. Right panel: Hopfield coefficients for each eigenenergy. There is a fifth trivial eigenvector; $\left|\lambda_{5} \right>=\left|G,0,G,0,1\right>$ with eigenenergy $\omega_{m}$ (Not shown here). Parameters: $\omega_{c1}=\omega_{0}-\Delta/2$, $\omega_{c2}=\omega_{0}+\Delta/2$, $g_{m1}=g_{m2}=g_{m}$, $\omega_{0}=20g_{m}$, $\omega_{a1}=20.1g_{m}$, $\omega_{a2}=19.9g_{m}$, $\omega_{m}=3g_{m}$ and $g_{1}=g_{2}=g_{m}/20$.}
\label{eigenenergies}       
\end{figure*}

One interesting feature to find out from the diagonalization of the Hamiltonian \ref{total_hamiltonian} is a region with anticrossing between both cavities which evidences a photonic molecule regime. By comparison of the dashed and color lines in Fig. \ref{eigenenergies} we can observe a blue shift of the energy of each quantum emitter caused by the mechanical resonator. This is an immediate effect of the atom-phonon dispersive coupling because of the significant detuning between both quantum emitters. As a consequence, the light-matter anticrossing in each emitter-cavity subsystem is changed and hence the polariton energies are also shifted without an appreciable modification of the energy splitting.


Two new and interesting types of anticrossing regions arise in the dispersive diagram. One of these is a high effective coupling between a quantum emitter and a cavity from different subsystems, Atom1-Cavity2 and Atom2-Cavity1 in Fig. \ref{eigenenergies}. It means reaching a light-matter strong coupling regime with atoms and cavities spatially separated and hence a high light-matter entanglement between distant subsystems. As shown in Fig. \ref{eigenenergies}, the eigenvectors at specific detuning values are mainly $\left|X,0,G,0,0\right>$ $\pm$ $\left|G,0,G,1,0\right>$ and $\left|G,0,X,0,0\right>\pm\left|G,1,G,0,0\right>$. Furthermore, as is derived below, this effective interaction rate is $\frac{g_{2} g_{m}^{2}}{\Delta_{m}\Delta_{a}^{21}}$ with $\Delta_{m}=\omega_{a2}-\omega_{m}$, $\Delta_{a}^{21}=\omega_{a2}-\omega_{a1}$ and $g_{m1}=g_{m2}=g_{m}$. Another attractive feature found at resonance in the energy spectrum and Hopfield coefficients ($\ket{\lambda_{2}}$ and $\ket{\lambda_{3}}$)  is the normal mode splitting between both cavities which leads to a highly all-photonic dressed state, i.e., a photonic molecule regime: $\ket{\psi}\approx \frac{1}{\sqrt{2}}(\ket{0,1,0,0,0}\pm\ket{0,0,0,1,0})$. The effective interaction rate in this case is $\frac{g^{2}g_{m}^{2}}{\Delta_{m}\Delta_{a}^{21}\Delta_{ac}}$ with $\Delta_{ac}=\omega_{a}-\omega_{c}$ and $g_{1}=g_{2}=g$. Upper in the states ladder, the dressing of the states is non-trivial and eigenstates are combination of almost all states involved in each excitation manifold. For this reason, the physical results in this paper are valid in a low excitation regime.\\

In order to analyze each anticrossing region, the Hamiltonian is rewritten in a bare part and an interaction Hamiltonian; $\hat{H}=\hat{H}_{0}+\hat{H}_{int}$, with 

\begin{eqnarray}
\hat{H}_{0}= \omega_{m}\hat{b}^{\dagger}\hat{b}+ \omega_{c1}\hat{a}_{1}^{\dagger}\hat{a}_{1}+\omega_{a1}\hat{\sigma}_{1}^{\dagger}\hat{\sigma}_{1}\nonumber \\
+\omega_{c2}\hat{a}_{2}^{\dagger}\hat{a}_{2}+\omega_{a2}\hat{\sigma}_{2}^{\dagger}\hat{\sigma}_{2} 
\end{eqnarray}
and
\begin{eqnarray}
\hat{H}_{int}=g_{1}\left(\hat{a}_{1}^{\dagger}\hat{\sigma}_{1}+\hat{a}_{1}\hat{\sigma}_{1}^{\dagger}\right)+g_{2}\left(\hat{a}_{2}^{\dagger}\hat{\sigma}_{2}+\hat{a}_{2}\hat{\sigma}_{2}^{\dagger}\right) \nonumber \\
+g_{m1}\left(\hat{b}^{\dagger}\hat{\sigma}_{1}+\hat{b}\hat{\sigma}_{1}^{\dagger} \right)+g_{m2}\left(\hat{b}^{\dagger}\hat{\sigma}_{2}+\hat{b}\hat{\sigma}_{2}^{\dagger} \right)
\end{eqnarray}

Now, we transform the Hamiltonian into the interaction picture, $\hat{H}_{IP}(t)=e^{i\hat{H}_{0}t}\hat{H}_{int}e^{-\hat{H}_{0}t}$:

\begin{eqnarray}
\hat{H}_{IP}=g_{1}\left(a_{1}^{\dagger}\sigma_{1}e^{i(\omega_{c1}-\omega_{a1})t}+a_{1}\sigma_{1}^{\dagger}e^{-i(\omega_{c1}-\omega_{a1})t} \right) \nonumber \\
+g_{2}\left(a_{2}^{\dagger}\sigma_{2}e^{i(\omega_{c2}-\omega_{a2})t}+a_{2}\sigma_{2}^{\dagger}e^{-i(\omega_{c2}-\omega_{a2})t} \right) \nonumber \\
+g_{m1}\left(\sigma_{1}^{\dagger}b e^{i(\omega_{a1}-\omega_{m})t}+\sigma_{1}b^{\dagger}e^{-i(\omega_{a1}-\omega_{m})t} \right) \nonumber \\
+g_{m2}\left(\sigma_{2}^{\dagger}b e^{i(\omega_{a2}-\omega_{m})t}+\sigma_{2}b^{\dagger}e^{-i(\omega_{a2}-\omega_{m})t} \right) 
\end{eqnarray}
Having this on mind, a formal integration of the Schr\"odinger equation is carried out, $\left|\Psi_{IP}(t)\right>=\mathcal{T}\left[e^{-i \int_{0}^{t} H_{IP}(t')dt'} \right]\left|\Psi_{IP}(0)\right>$, in order to perform the approximation of large detuning between the atom and the mechanical resonator, $\omega_{a2}\gg \omega_{m}$ and $\omega_{a1}\gg \omega_{m}$. The propagator can be expressed as a perturbation expansion:

\begin{eqnarray}
\mathcal{T}\left[e^{-i \int_{0}^{t} H_{IP}(t')dt'} \right]&=&\hat{1}-i\int_{0}^{t}\hat{H}_{IP}dt' + \mathcal{O}^{2}(H_{IP}) +\cdots \nonumber \\
&\approx & \hat{1}-i \hat{H}_{eff}t
\end{eqnarray}

Only the first four terms of the series contribute to the effective Hamiltonian:\\

\textbf{First order} 

\begin{equation}
\hat{H}_{eff}^{(1)}=g_{1}\left(a_{1}^{\dagger}\sigma_{1}+a_{1}\sigma_{1}^{\dagger} \right)+g_{2}\left(a_{2}^{\dagger}\sigma_{2}+a_{2}\sigma_{2}^{\dagger} \right)
\end{equation}

\textbf{Second order}

Assuming $\omega_{a2}\approx\omega_{a1} $

\begin{eqnarray}
\hat{H}_{eff}^{(2)}&=&\frac{g_{m1}^{2}}{\Delta_{m}}\left( \sigma_{1}^{\dagger}\sigma_{1}+b^{\dagger}b\sigma_{z1}\right)+\frac{g_{m2}^{2}}{\Delta_{m}}\left( \sigma_{2}^{\dagger}\sigma_{2}+b^{\dagger}b\sigma_{z2}\right) \nonumber \\
&+&\frac{g_{m1}g_{m2}}{\Delta_{m}}\left(\sigma_{1}^{\dagger}\sigma_{2} + \sigma_{1}\sigma_{2}^{\dagger} \right)
\end{eqnarray}

\textbf{Third order}

Assuming $\omega_{a2}\approx\omega_{c1}$ and $\omega_{a1}\approx\omega_{c2}$, but $\omega_{a1}\neq\omega_{a2}$

\begin{eqnarray}
\hat{H}_{eff}^{(3)}=\frac{g_{1}g_{m1}g_{m2}}{\Delta_{m}\Delta_{a}^{21}}\sigma_{1}\sigma_{1}^{\dagger}\left(a_{1}^{\dagger}\sigma_{2}+a_{1}\sigma_{2}^{\dagger}\right) \nonumber \\
+\frac{g_{2}g_{m1}g_{m2}}{\Delta_{m}\Delta_{a}^{12}}\sigma_{2}\sigma_{2}^{\dagger}\left(a_{2}^{\dagger}\sigma_{1}+a_{2}\sigma_{1}^{\dagger}\right) 
\end{eqnarray}

\textbf{Fourth order} 

Assuming $\omega_{c1}\approx\omega_{c2}$

\begin{eqnarray}
\hat{H}_{eff}^{(4)}=\frac{g_{1}g_{2}g_{m1}g_{m2}}{\Delta_{m}\Delta_{a}^{21}\Delta_{ac}^{21}}\sigma_{1}^{\dagger}\sigma_{1}\sigma_{2}^{\dagger}\sigma_{2}(a_{1}a_{2}^{\dagger}+a_{1}^{\dagger}a_{2}) \nonumber \\
+\frac{g_{1}g_{2}g_{m1}g_{m2}}{\Delta_{m}\Delta_{a}^{12}\Delta_{ac}^{12}}\sigma_{1}^{\dagger}\sigma_{1}\sigma_{2}^{\dagger}\sigma_{2}(a_{1}a_{2}^{\dagger}+a_{1}^{\dagger}a_{2})
\end{eqnarray}

with $\Delta_{a}^{ij}=\omega_{ai}-\omega_{aj}$ and $\Delta_{ac}^{ij}=\omega_{ai}-\omega_{cj}$.

\begin{figure}[H]
\centering
\resizebox{0.45\textwidth}{!}{%
\includegraphics{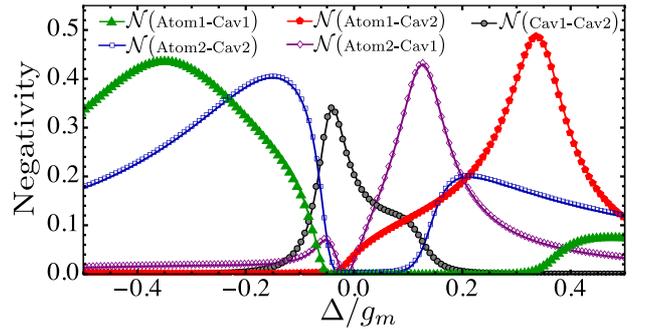}
}
\caption{Bipartite entanglement. Negativity between both cavities (Black line), between an atom and a cavity from the same subsystem (Blue and green lines) and between and atom and a cavity from different subsystems (Purple and red lines). Parameters: $g_{m1}=g_{m2}=g_{m}$, $\omega_{0}=20g_{m}$, $\omega_{a1}=20.1g_{m}$, $\omega_{a2}=19.9g_{m}$, $\omega_{m}=3g_{m}$ and $g_{1}=g_{2}=g_{m}/20$.}
\label{entanglement}
\end{figure}

\subsection{Entanglement properties}

The next natural step is to analyze the entanglement properties of each anticrossing. Negativity between two parts of the system is computed by tracing over the other degrees of freedom, i.e. $\mathcal{N}=\sum_{\lambda<0}|\lambda|$, where $\lambda$ denotes all the eigenvalues of the partial transpose of the traced density matrix $\rho_{B}=Tr_{A}(\rho_{AB})$. $B$ represents the two parts of interest and $A$, the other parts. Here $\rho_{AB}$ is the pure density matrix build with the eigenstates, $\ket{\lambda_{i}}$, of the Hamiltonian \ref{total_hamiltonian}. The eigenstate used is the one involved in the anticrossing of interest, e.g., entanglement between atom 1 and cavity 2 is computed with the eigenstate $\ket{\lambda_{4}}$ shown in Fig. \ref{eigenenergies}.

As shown in Fig. \ref{entanglement}, the largest entanglement between photons from both cavities is found close to the resonance; $\omega_{c1}\approx\omega_{c2}$ as expected according to the Fig. \ref{eigenenergies}. Furthermore, direct light-matter entanglement is maximum in $\omega_{a1}\approx\omega_{c1}$ and $\omega_{a2}\approx\omega_{c2}$, and indirect light-matter entanglement increases around $\omega_{a2}\approx\omega_{c1}$ and $\omega_{a1}\approx\omega_{c2}$, also expected from the analysis made in the previous section. The above means that the best condition to entangle two parts of the system is setting those parts close to resonance. However, due to additional contributions in the eigenstates (Fig. \ref{eigenenergies}), there is a coexistence of the different types of entanglement studied, e.g., at the maximum of entanglement between atom 1 and cavity 2, there is also entanglement atom2-cav2 and atom2-cav1. Additionally, the phonon part of the system is unentangled with the rest of the system as long as the mechanical frequencies are much smaller than the photon ones.

\begin{figure}[H]
\centering
\resizebox{0.47\textwidth}{!}{%
\includegraphics{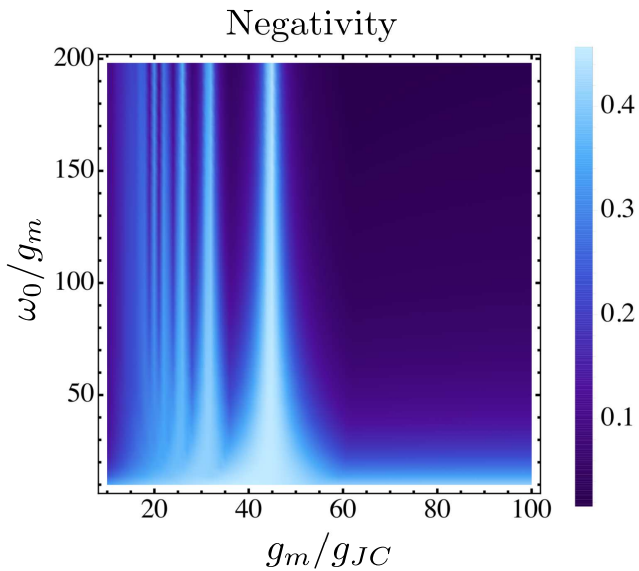}
}
\caption{Photon-photon entanglement. Maximum of the negativity of the reduced cavity-cavity system as a function of $\omega_{0}/g_{m}$ and $g_{m}/g_{JC}$. Parameters: $g_{1}=g_{2}=g_{JC}$, $g_{m1}=g_{m2}=g_{m}$, $\omega_{a1}=\omega_{0}+0.1$, $\omega_{a2}=\omega_{0}-0.1$, $\omega_{c1}=\omega_{0}-\Delta/2$, $\omega_{c2}=\omega_{0}+\Delta/2$ and $\omega_{m}=3g_{m}$.}
\label{entanglement2}
\end{figure}

Now, in order to study the dependence of the entanglement with the physical parameters, we focus on the intercavity photon entanglement. There are two critical parameters in the system; which are the ratio between the atom/cavity frequencies or the light-matter coupling strength with the mechanical interaction rate, $\omega_{0}/g_{m}$ and $g_{JC}/g_{m}$, respectively. Fig. \ref{entanglement2} exhibits fringes of maximum entanglement for mechanical interaction rates larger than the light-matter ones, i.e., when the coupling strength between both subsystems exceed the coupling strength in each subsystem. This fringes show that entanglement does not change for large ratios $\omega_{0}/g_{m}$, which is the most typical condition in current experimental setups, especially in circuit QED systems. For small ratios between photon frequencies and mechanical interaction, $\omega_{0}/g_{m}<30$, the intercavity photon entanglement is less sensitive to the change of $g_{m}/g_{JC}$. Nevertheless, in order to keep the rotating wave approximation for light-matter interaction, it is necessary to fulfill the condition $\omega_{0}\gg g_{JC}$, i.e., $g_{m}>g_{JC}$ in our model. In other words, the system must fulfill that the atom/cavity frequencies exceed the mechanical interaction rate and, subsequently, this last one should exceed the light-matter coupling strength. Finally, in all calculations performed it was keep the condition $\omega_{m}=3g_{m}$, which is possible to reach in nowadays experiments.

\section{Discussion and Conclusions}
\label{conclusions}

In this work, we have considered a single mode of a mechanical resonator mediating the interaction between two subsystems, each one composed of a quantum emitter coupled to a cavity. As we have seen, it is possible to reach a regime of parameters with normal mode splitting between the cavities which is a first signature of photonic molecules. Besides, it was found Vacuum Rabi splitting between a quantum emitter and a cavity from different subsystems, i.e., non-local light-matter strong coupling regime. As a direct effect of the mechanical resonator, it was observed a blue shift of the atomic energies and hence a shift in the polariton energies. Additionally, it was studied the entanglement properties of those dressed states and the conditions of the physical parameters for which it is maximized the entanglement. The multipartite system evidences that some parts of the system are most entangled than others depending on the energy detuning or, more specifically, on the dressed state involved. Good candidates to implement our proposal are circuit quantum electrodynamics systems where high mechanical interactions could be reached. 

Finally, if the mechanical resonator couples the cavities instead of the quantum emitters, same results are found since at first excitation manifold both situations are equivalents. However, beyond the low excitation regime, where higher excitation manifolds are involved, the dressing of the states and the entanglement change substantially due to the statistics of the particles involved in the mechanical interaction; two-level systems (artificial atoms) or bosons (cavities). 

\section*{Acknowledgements}
The authors acknowledge partial financial support from COLCIENCIAS under the project ``Emisi\'on en sistemas de Qubits Superconductores acoplados a la radiaci\'on. C\'o-\break digo 110171249692, CT 293-2016, HERMES 31361" and the project ``Optomec\'anica y electrodin\'amica con puntos cu\'anticos en microcavidades. C\'odigo 201010027618, HERMES 39177 ". J.E.R. thanks financial support from the ``Beca de Doctorados Nacionales de COLCIENCIAS 727" and J.P.R.C. is grateful to the ``Beca de Doctorados Nacionales de COLCIENCIAS 785". 

\section*{Authors contributions}
All the authors were involved in the preparation of the manuscript.
All the authors have read and approved the final manuscript.
%

 \bibliographystyle{unsrt}
 \bibliography{Ref.bib}

\end{document}